\documentclass[%
 aip,
 amsmath,amssymb,
 reprint,%
]{revtex4-1}

\usepackage{graphicx}% Include figure files
\usepackage{dcolumn}% Align table columns on decimal point
\usepackage{bm}% bold math
\usepackage[mathlines]{lineno}% Enable numbering of text and display math
\usepackage{xcolor}
\usepackage[utf8]{inputenc}
\usepackage[T1]{fontenc}
\usepackage{mathptmx}
\usepackage{etoolbox}
\graphicspath{{images/}}
\usepackage{float}
\usepackage{mathtools,amsmath, amssymb,amsthm} 
\usepackage{multirow,tabularx,makecell}
\usepackage[thinlines]{easytable}
\newcolumntype{Y}{>{\centering\arraybackslash}X}
\usepackage{url}
\usepackage[pdftex,colorlinks]{hyperref}
\definecolor{darkblue}{cmyk}{1,1,0,0}
\definecolor{darkred}{cmyk}{0,1,0,0.7}
\definecolor{darkgrey}{cmyk}{0,0,0,0.8}
\hypersetup{anchorcolor=.,
  citecolor=., filecolor=.,
  menucolor=.,pagecolor=.,urlcolor=.,linkcolor=.}
\DeclareMathOperator{\sig}{S}

\newcommand{\rp}{{r_\textrm{p}}}
\newcommand{\df}{{d_\textrm{f}}}
\newcommand{\Tinp}{{T_\textrm{inp}}}
\newcommand{\thi}{{\theta_\mathrm{in}}}
\newcommand{\tho}{{\theta}}

\definecolor{darkgreen}{RGB}{0,127,0}

\renewcommand{\arraystretch}{1.2}

\makeatletter
\def\@email#1#2{%
 \endgroup
 \patchcmd{\titleblock@produce}
  {\frontmatter@RRAPformat}
  {\frontmatter@RRAPformat{\produce@RRAP{*#1\href{mailto:#2}{#2}}}\frontmatter@RRAPformat}
  {}{}
}%
\makeatother
\begin{document}

\preprint{AIP/123-QED}

\title{Predicting Perceptual Boundaries in Auditory Streaming using Delay Differential Equations}
% Force line breaks with \\
\author{Asim Alawfi}
\email{aa1045@exeter.ac.uk}
%\affiliation{Department of Mathematics and Statistics, University of Exeter, Harrison Building, Exeter EX4 4QF, UK.}
\affiliation{ 
Department of Mathematics and Statistics, College of Science, Imam Mohammad Ibn Saud Islamic University (IMSIU), Riyadh 11566, Saudi Arabia.
}
\author{Farzaneh Darki}%
%\email{f.darki2@exeter.ac.uk}
%\affiliation{ 
%Authors' institution and/or address%\\This line break forced with \textbackslash\textbackslash
%}%
\author{Jan Sieber}
%\email{j.sieber@exeter.ac.uk}
\affiliation{Department of Mathematics and Statistics, University of Exeter, Harrison Building, Exeter EX4 4QF, UK.}

\date{\today}% It is always \today, today,

\begin{abstract}
Auditory streaming enables the brain to organize sequences of sounds into perceptually distinct sources, such as following a conversation in a noisy environment. A typical experiment for investigating perceptual boundaries and bistability is to present a subject with a stream containing two alternating tone stimuli. We investigate a model for the processing of such a stream consisting of two identical neural populations of excitatory and inhibitory neurons. The populations are coupled via delayed cross-inhibition and periodically forced with sharp step-type signals (the two-tone stream). 
We track how the perception boundary depends on threshold selection and establish how boundaries between three different auditory perceptions (single tone versus two tones versus bistability between both perceptions) relate to bifurcations such as symmetry breaking. We demonstrate that these transitions are governed by symmetry-breaking bifurcations and that the perceptual classification based on neural thresholds is highly sensitive to threshold choice. Our analysis reveals that a fixed threshold is insufficient to capture the true perceptual boundaries and proposes a variable-threshold criterion, informed by the amplitude dynamics of neural responses. Finally, we illustrate how key stimulus parameters such as tone duration,  delay,  and internal time scale shape the boundaries of auditory perceptual organization in the plane of the two most commonly varied experimental parameters, the representation rate, and the difference in tone frequency. These findings offer mechanistic insight into auditory perception dynamics and provide a refined framework for linking neural activity to perceptual organization.
\end{abstract}
\maketitle

%\section{\label{sec:level1}First-level heading:\protect\\ The line
%break was forced \lowercase{via} \textbackslash\textbackslash}
\begin{quotation}
   A research question in studies of auditory streaming is how our brain processes different sounds simultaneously. Experiments investigate this by presenting to listeners a sequence of sounds consisting of two different tones separated by a controlled difference in frequency ($\df$) and repeated at a controlled presentation rate ($\rp$). The experiments show two significant perceptual boundaries in $(\rp, \df)$-space: the coherence and the fission curve. Listeners report that they hear a single sound (called integrated) when presented tones below the fission curve (small $\df$, small $\rp$). For tones above the coherence curve (large $\df$, large $\rp$) they report hearing two separate sounds simultaneously (called segregated). For tones between these two curves they report that their perception alternates spontaneously between the integration and the segregation (called bistability). %Recently, \citet{J} proposed a model using delay differential equations that generates an approximation of the coherence and fission curves by counting the number of crossings of a fixed threshold by excitations of two identical neural populations. 
   
   We link these perceptual boundaries to a symmetry breaking in a mathematical model describing delayed coupling between the neuron populations for processing each tone. By tracking where symmetry breaks in the model we study the dependence of the perception boundary on neuron population properties that cannot be directly observed (such as activation thresholds and coupling delays). Our diagrams of perceptual boundaries in the $(\rp,\df)$-plane correspond to what is reported in experiments such that they can be used to infer these unobservable neuron population properties from experimental observations. 
\end{quotation}
\section{Introduction}
Auditory streaming in the brain organizes complex acoustic environments into perceptually distinct streams~\cite{bregman1994auditory}. This process enables one to perform tasks such as following a conversation in a crowded room~\cite{mcdermott2009cocktail}, focusing on a specific sound source while filtering out others, and distinguishing musical patterns~\cite{fishman2001neural}. Experimental~\cite{van1977minimum,13} and theoretical~\cite{46} studies on auditory streaming consider a stimulus consisting of a sequence of alternating high (which we label ``A'') and low (which we label ``B'') frequency tones. This sequence can be perceived either as one integrated sound (A\,B\,A\,B\,A\,B\,A\,B) or as two separate streams, with one stream representing tone A and the other representing tone B (concurrent: A\,-\,A\,-\,A\,-\,A\,- and -\,B\,-\,B\,-\,B\,-\,B, where each dash (``-”) represents a silence of tone duration)~\cite{moore2012properties, snyder2007toward}. Furthermore, listeners may experience spontaneous perceptual switches between these interpretations, a phenomenon known as perceptual bistability~\cite{denham2013perceptual}.\\

\begin{figure*}[htbp]
    \centering
        \includegraphics[width=\textwidth]{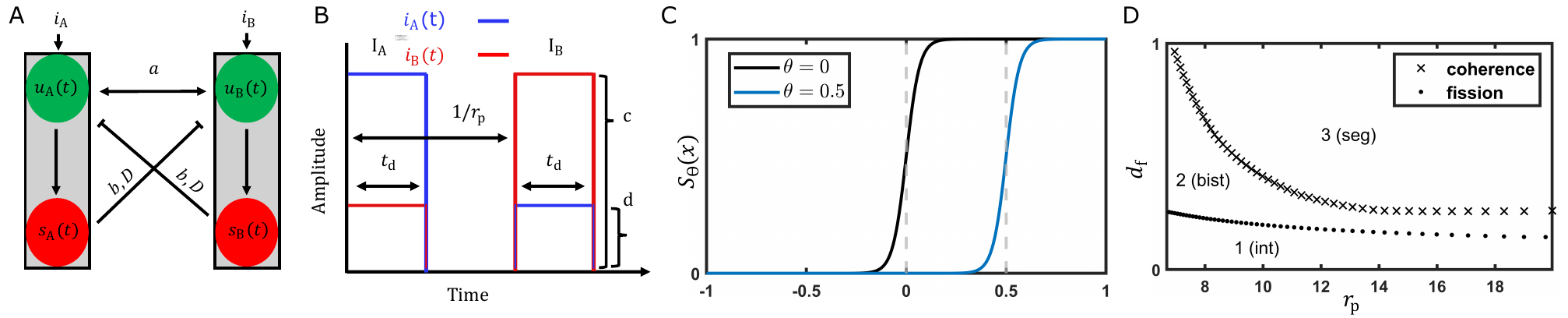}
    \caption{{\bf (A)} Block diagram showing excitatory and inhibitory connections between neural populations for tones A and B.  Inhibition is delayed by $D$. The inputs for the populations  $i_A$ and $i_B$ receive inputs have a periodic time profile shown in (B). {\bf (B)} Time profiles of inputs $i_A(t)$ and $i_B(t)$ over one period with labels for parameters presentation rate $r_{p}$, tone duration $t_{d}$, active tone amplitude $c$, and non-active tone amplitude $d$.The intervals during which tones A and B are active are denoted ${I_{A}}$ and ${I_{B}}$, respectively. Square-wave inputs $i_{A}(t)$ and $i_{B}(t)$ correspond to populations associated with tones A and B. {\bf (C)} Graph of Sigmoid function with slope $\lambda=30$ for $\thi=0$ and $\thi=0.5$.  {\bf (D)} Experimentally detected fission and coherence boundaries (original data from \citet{46}, digitized\citep{J}).} 
    \label{Sigmoid_plot}
\end{figure*}

When presenting an alternating two-tone stream to a listener in an experiment the frequency difference $\df$ between the two tones and the presentation rate $\rp$ are the two main controllable stimulus parameters.\cite{Van} The pioneering experimental work in 1975 by \citet{Van} mapped  perceptual outcomes in the $(\df,\rp)$-plane to obtain the van Norden diagram, shown in Figure~\ref{Sigmoid_plot}C. The study found that the $(\df,\rp)$-plane is divided by two critical curves, the fission and coherence boundaries. Below the fission boundary, listeners consistently perceived the sequence as integrated~\cite{elhilali2009temporal}. Above the coherence boundary, the perception was predominantly segregated~\cite{van1977minimum}. In the intermediate region between these boundaries, listeners reported perceptual bistability~\cite{denham2013perceptual}.\\

Higher-level processing of acoustic stimuli from the environment is performed in the auditory cortex. The primary auditory cortex is tonotopically organized, meaning that neurons are arranged to respond selectively to specific frequency ranges in a gradient from low to high frequencies~\cite{humphries2010tonotopic}. The secondary auditory cortex receives inputs from the primary auditory cortex~\cite{wang2007neural}. Cross-inhibition between neuronal populations  in the secondary auditory cortex tuned to different frequencies supports frequency selection\cite{kato2017network, natan2017cortical}. Several studies have explored the neural mechanisms underlying auditory streaming, particularly how neural populations encode and differentiate between perceptual outcomes such as integration, segregation, and bistability~\cite{snyder2017recent}. Fishman et al.~\cite{13} investigated how factors such as frequency difference $\df$, tone duration $t_{d}$, and presentation rate $\rp$ influence stream segregation in the primary auditory cortex of macaque monkeys. They found that increasing $\df$ or $\rp$ enhances neural selectivity, thereby facilitating perceptual segregation (agreeing qualitatively with the van Norden diagram). However, their study was limited in scope, exploring only discrete values of $\rp$ and $\df$, without offering an explanation for nonlinear dynamical phenomena such as bistability between perceptual states. 

Theoretical research addresses these gaps by modelling the interactions between excitatory and inhibitory neural populations, which are thought to control the dynamics of perceptual bistability~\cite{rankin2019computational}. \citet{46} considered a neural field description with a continuous representation of tonotopy to simulate the dynamics of auditory streaming. \citet{rankin2015neuromechanistic} considered a model based on a discrete idealization of a tonotopically organized array, incorporating recurrent excitation, mutual inhibition, and slow processes such as adaptation~\cite{rankin2017stimulus,byrne2019auditory,rankin2022attentional}. \citet{J} developed a simpler two-population model that includes fast excitation and slow delayed inhibition. All of these models estimate key features of auditory perceptual organization, as described by the van Noorden diagram \citep{Van}. Bistability is a common feature of these models, enabling transitions between different perceptual states. Sometimes bistability is inherent in the processing of the neural activity, as seen in stream competition models~\cite{rankin2015neuromechanistic,J}, while in other cases, it arises through neural activation in a different, non-tonotopically organized cortical area (stream classification models)~\cite{46}.\\

Different approaches exist to linking neural activities to perceptual phenomena in auditory streaming models. Perceptual classifications are often derived through numerical simulations based on the dominance of specific neural populations (winner-take-all, WTA)~\cite{rankin2015neuromechanistic} or the number of threshold crossings in neural population activity during a single periodic interval of stimulus~\cite{J}. In this study, we consider the auditory streaming model presented by~\citet{J}, focusing on its ability to classify neural activity into three perceptual outcomes: integration, segregation, and bistability. While the model successfully identifies a parameter region with bistable dynamical behavior, we observe a mismatch between the boundary of perceptual classification and the region of bistable dynamics.

Our goal is to provide a more detailed classification of perceptual states in $(\rp, \df)$-space in the Ferrario\,\&\,Rankin model, to extend the findings of \citet{J,ferrario2021cascades}, and to examine how additional parameters such as tone duration $t_d$ and delay $D$ influence perceptual boundaries in $(\rp, \df)$-space. Building on the auditory model proposed by \citet{J}, we reformulate the periodically forced model into an extended autonomous system to facilitate bifurcation analysis using \texttt{DDE-Biftool}. We also investigate the role of threshold variability in perceptual classification. By computing the amplitude of neural activity across the bistable dynamical region, we find that the threshold for classification is not fixed. Specifically, the threshold is lower for transitions from segregation to bistability and higher for transitions from integration to bistability. Additionally, we explore the role of a faster internal time scale $\tau$ to better understand its impact on auditory streaming and perceptual organization.
\section{Mathematical model for auditory streaming system}\label{sec:audi:model}
\subsection{Neural mass model of secondary auditory cortex}\label{sec_original_model}
We use a model for processing of auditory perception when a listener is presented with a sequence consisting of stimuli alternating between tone A and tone B in the form $(\mathrm{ABAB}\ldots)$. The tones have a difference in frequency $\df$ and alternate with a presentation rate $\rp$.\\
\paragraph{Differential equations for coupled neural populations.}
Originally developed by \citet{J}, the model couples two identical neuron populations, each responding to one of the two tones, and each with a cross-excitatory and a cross-inhibitory component. Figure~\ref{Sigmoid_plot}A shows a block diagram with the connections between the two involved neural populations. We model  the cross-inhibition as delayed with a discrete delay $D$, such that the block diagram in Figure~\ref{Sigmoid_plot}A is described by a four-dimensional system of  delay differential equations with periodic forcing: 
\begin{equation} \label{model}
\begin{split}
\tau\dot{u}_A(t)&=-u_A(t)+\sig_{\thi}\left(au_B(t)-bs_B(t-D)+i_A(t)\right), \\
\tau\dot{u}_B(t)&=-u_B(t)+\sig_{\thi}\left(au_A(t)-bs_A(t-D)+i_B(t)\right),\\
\tau\dot{s}_A(t)&=\sig_{\thi}(u_A(t))\left(1-s_A(t)\right) -\frac{\tau}{\tau_{i}} s_A(t), \\
\tau\dot{s}_B(t)&=\sig_{\thi}(u_B(t))\left(1-s_B(t)\right) - \frac{\tau}{\tau_{i}} s_B(t). 
\end{split}
\end{equation}
In \eqref{model} $u_{A}$ and $u_{B}$ are the average firing rate for the population of excitatory neurons responding to tone A and tone B, respectively, with timescale $\tau$. Mutual coupling through direct fast excitation has strength $a$. Firing of the inhibitory neural populations, described by the variables $s_{A}$ and $s_{B}$, cross-inhibits the neural populations $u_{A}$ and $u_{B}$ with strength  $b$, timescale $\tau_i$ (where $\tau \ll \tau_{i}$), and delay $D$. The coupling is passed through the Sigmoid activation function $\sig_\thi$ (see Figure~\ref{Sigmoid_plot}C) with threshold $\thi$ and maximal slope $\lambda/4\gg1$ at $\thi$, such that $\sig_{\theta}(x)=\left[1+\exp(-\lambda (x-\theta))\right]^{-1}$.

\begin{table}
\caption{\label{tab:parameters:exper} Parameters of model \eqref{model},\eqref{inpts_smooth_fun} and \eqref{modelhopf}. Definitions and values from experimental data reported in~\citet{J}.}
\begin{ruledtabular}
\begin{tabular}{lcr}
Parameter& Definition& Values
\\
\hline\\[-2.5ex]
$\rp$ & presentation rate & $[1,40]$\,Hz\\
$\df$&  tone frequency difference  & $\df\in [0,1]$    \\
$D$ & delay & 0.015\,s  \\
$t_{d}$ & tone duration & $0.022$\,s  \\ 
$\tau_i$ & external time scale & $0.25$\,s \\ 
$\tau$ & internal time scale & $0.025$\,s  \\
$a$ & strength of fast cross excitation & 2  \\ 
$b$ & strength of cross inhibition& 2.8  \\ 
$c$ & tone amplitude & 5.5 \\ 
$m$ & frequency scaling factor & 6 \\
$\lambda$ & slope parameter of Sigmoid activation $\sig_\theta$& $30$\\
$\thi$ & threshold of Sigmoid activation $\sig_{\thi}$& $0.5$\\
$\tho$ & threshold for perception detection& $0.5$\\
$\Tinp$ & forcing period & $2/\rp$\\
$\omega$ & frequency of harmonic oscillator in \eqref{modelhopf}& $\pi\rp$\\
$\alpha$ & neg.~damping of harmonic oscillator in \eqref{modelhopf} & $1$
\end{tabular}
\end{ruledtabular}
\end{table}
Slow inhibition masks the perception of subsequent tones during segregation (\emph{forward masking}~\cite{moore2012introduction}), whereas fast excitation allows integration of large pitch differences between the two tones. Inhibition can be influenced by factors such as slower activation times of inhibitory pathways compared to excitatory ones~\cite{park2020circuit} (controlled by the ratio $\tau_i/\tau$) indirect connections via interneurons~\cite{moore2013parvalbumin}, and propagation delays between spatially separated populations A and B~\cite{reimer2011fast} (controlled by discrete delay $D$).

\paragraph{Input from primary auditory cortex.}
Each neuron in the primary auditory cortex exhibits a characteristic frequency, referred to as its best frequency (BF), at which it demonstrates peak responsiveness~\cite{sutter1999organization}. However, these neurons also respond to a lesser degree to frequencies outside their BF, termed non-best frequencies~\cite{schreiner2007auditory}. The differential activity of neurons, characterized by robust responses to BFs and attenuated responses to non-BFs, underpins the brain's capacity to organize complex auditory stimuli into perceptually distinct streams\cite{micheyl2005perceptual, riecke2018frequency}. For example, neurons tuned to the frequencies of tones A and B facilitate auditory stream segregation by amplifying differences in neural activity patterns when the tones differ significantly in frequency~\cite{bee2004primitive}. Conversely, when the frequencies are similar, overlapping neural responses contribute to perceptual integration~\cite{13}.

Figure~\ref{Sigmoid_plot}B shows the inputs (forcing) $i_{A}(t)$ and $i_{B}(t)$. They are square-wave functions that model the output from neurons in the primary auditory cortex, which project into the secondary auditory cortex, when listening to the sequence of A and B tones. The periodic inputs $\left(i_{A}, i_{B}\right)$ generate a sequence of non-overlapping intervals called active tone intervals (labeled as ${I_{A}}$ and ${I_{B}}$ with ${I_{A}}\cap {I_{B}}=\emptyset$ in Figure~\ref{Sigmoid_plot}B). During each active interval inputs to both populations are non-zero. For example, during active tone interval $I_A$, the input $i_A$ has amplitude $c$, while the input $i_B$ has the smaller amplitude $d$, while the amplitudes are reversed during $I_B$. The ratio between the amplitude $d$ and the amplitude $c$ depends on the effective frequency difference $\df$ between the auditory stimuli according to the relation \begin{align*}
    d/c&=1-\df^{1/m},\mbox{\quad or\quad}
    \df=\left[\frac{c-d}{c}\right]^m.
\end{align*}
In this notation, the effective frequency difference $\df $ is a dimensionless parameter in the range $[0,1]$, which can be converted into semitone units using the formula $12 \log(1 + \df)$ according to \citet{J}. These show that the activity at tonotopic locations for B
decreases nonlinearly with $\df$ during A tone presentations. We construct the square-wave inputs $i_A$ and $i_B$ using the Sigmoid with a large slope as
\begin{align}
    \begin{split}
        i_{A}(t)=&\ c \sig_{0}(\sin(\pi \rp\ t))\sig_{0}(-\sin(\pi \rp\ (t-t_{d})))+  \\ &\ d \sig_{0}(-\sin(\pi \rp\ t))\sig_{0}(\sin(\pi \rp\ (t-t_{d}))),\\
        i_{B}(t)=&\ d \sig_{0}(\sin(\pi \rp\ t))\sig_{0}(-\sin(\pi \rp\ (t-t_{d})))+ \\ &\  c \sig_{0}(-\sin(\pi \rp\ t))\sig_{0}(\sin(\pi \rp\ (t-t_{d}))),
    \end{split} \label{inpts_smooth_fun}
\end{align}
where  $\sig_{0}$ is the Sigmoid function with $\theta=0$. 
%dominates on ${I_{A}}$, and the response to tone B dominates on  ${I_{B}}$. 
The length of each active tone interval is $t_\text{d}$ (equal for ${I_{A}}$ and ${I_{B}}$). We call this time  \emph{tone duration}. This is not a parameter controlled by the experimenter because it corresponds to the period of the activity of the primary auditory cortex in response to the experimental signal. Presentation rate $\rp$ is the rate at which active tone intervals are presented, such that the overall period $\Tinp$ of the forcing needs to satisfy 
\begin{align}\label{Tinp_rp}
    \Tinp=2/\rp>2t_\mathrm{d},    
\end{align}
to avoid overlapping of successive active tone intervals $I_A$ and $I_B$. The factor $2$ accounts for the two alternating tones presented in each forcing interval, as shown in Fig.~\ref{Sigmoid_plot}B. For our fixed $t_d$ from Table~\ref{tab:parameters:exper}, this restricts $\rp$ to the indicated range of up to $40$\,Hz.

\begin{figure}[htbp]
\centering
\includegraphics[width=0.45\textwidth]{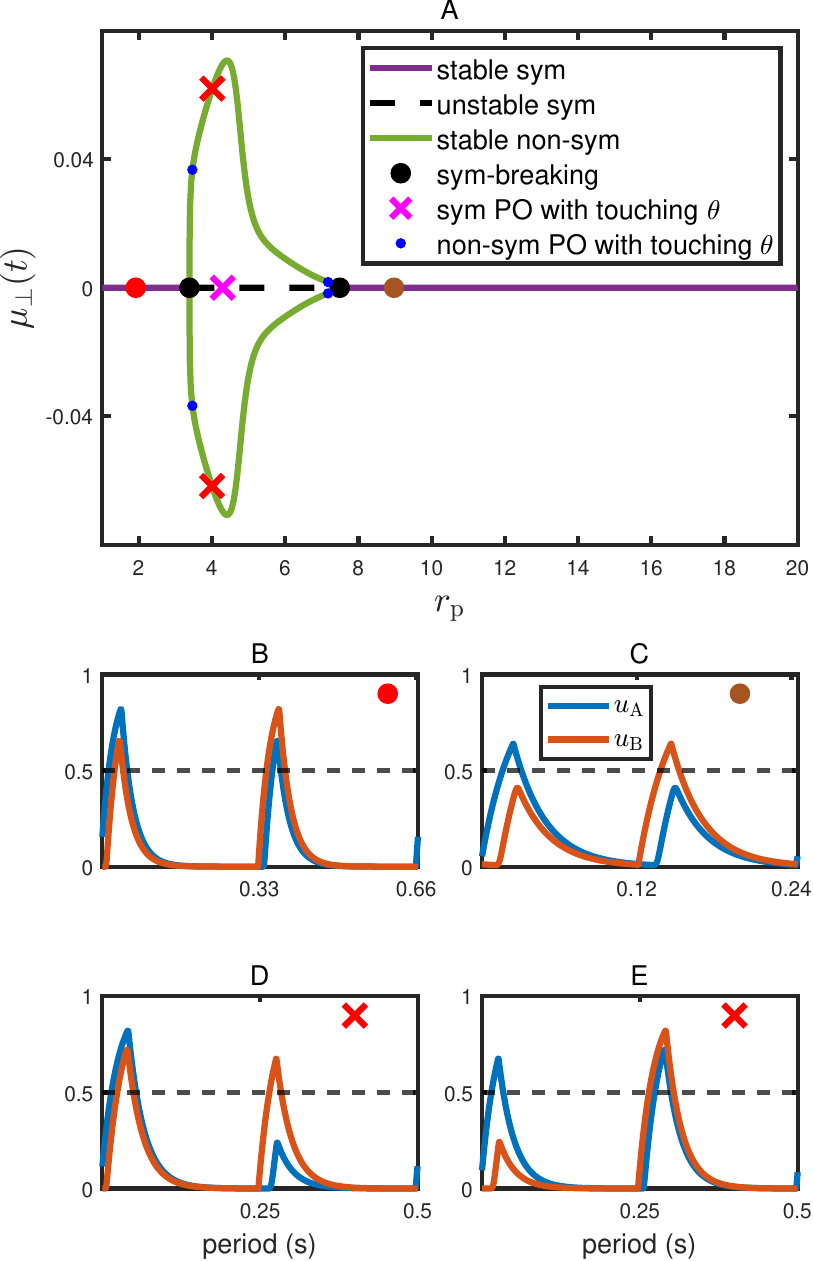}
\caption{{\bf (A)} The one-parameter bifurcation analysis of periodic orbits and their stability for the auditory streaming model \eqref{modelhopf} under the variation of $\rp$, while keeping the frequency difference fixed at $\df=0.73$. The violet solid line and black dashed line represent stable and unstable symmetric POs, respectively, while the green solid line represents the stable nonsymmetric POs. The $y$-axis represents the symmetry measure $\mu$ defined in~\eqref{measure_symmetry}. {\bf (B-C)} Time profiles of the POs highlighted by red and brown dots in panel~A, representing the integrated and segregated regions along the violet branch before and after the stability change. {\bf (D-E)} Time profiles of the stable nonsymmetric POs, marked with red crosses in panel~A, illustrating the bistability solutions of the model. All other parameters are as specified in Table~\ref{tab:parameters:exper}. 
}\label{po_pr_with_df}
\end{figure}

\begin{figure*}
    \centering
    \includegraphics[width=\textwidth]{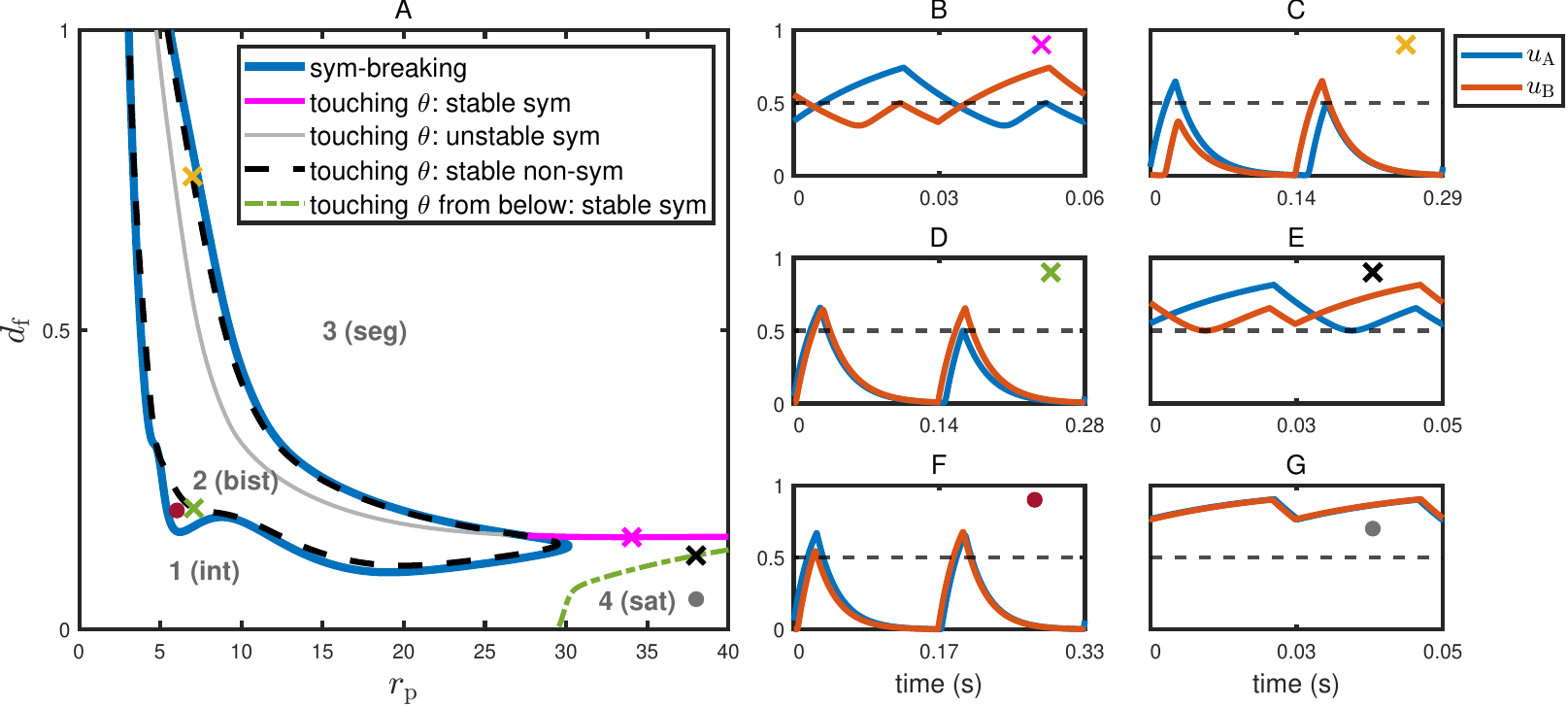}
    \caption{{\bf (A)} Two-parameter bifurcation diagram in the $(\rp,\df)$-plan. The blue curve represents the symmetry-breaking bifurcation. The pink and grey branches correspond to stable and unstable symmetric periodic orbit solutions that touch the threshold, respectively. The black dashed line represents the stable nonsymmetric periodic solutions that touch the threshold. The dashed green line corresponds to stable symmetric periodic orbits that touch the threshold from above. Regions 1, 2 and 3 correspond to auditory perceptual regions for integrated, bistability, and segregated, while region 4 corresponds to saturated behavior. {\bf (B)--(F)} Time profiles of periodic solutions indicated by color markers in panel~A. All other parameters are as specified in Table~\ref{tab:parameters:exper}.}
    \label{fig:two_parameter_orgcase2}
  \end{figure*}
\paragraph{Permutation symmetry} The inputs and the model have a $\mathbb{Z}_2$ symmetry $P$
\begin{multline}
  \label{eq:symmetry}
  P(u_A(\cdot),u_B(\cdot),s_A(\cdot),s_B(\cdot))=
  (u_B(\Tinp/2+(\cdot)),\\ u_A(\Tinp/2+(\cdot)),
  s_B(\Tinp/2+(\cdot)),s_A(\Tinp/2+(\cdot)),
\end{multline}
such that we expect to see \emph{symmetric} time profiles, where $u_A(t)=u_B(t+\Tinp/2)$, $u_B(t)=u_A(t+\Tinp/2)$, $s_A(t)=s_B(t+\Tinp/2)$, $s_B(t)=s_A(t+\Tinp/2)$ for all $t$. For any non-symmetric time profile we expect to see its symmetric counterpart with the same stability.
\subsection{Perception classification} 
Responses of model~\eqref{model} for different parameters $\df$ and $\rp$ are classified as \emph{integrated}, \emph{segregated}, or \emph{bistable} based on the total number of crossings of a threshold
\begin{align*}
    \tho=0.5
\end{align*} by $u_A(t)$ and $u_B(t)$ during one period of forcing shown in Figure~\ref{Sigmoid_plot}B with the two active tone intervals $I_A$ and $I_B$. We associate the \emph{integrated} percept (as reported by listeners in experiments) with the response where the threshold  $\tho$ is crossed overall four times: twice by  $u_A(t)$ and twice by $u_B(t)$, as shown in the example time profile Figure~\ref{po_pr_with_df}B. Both, $u_A(t)$ (blue) and $u_B(t)$ (red) go above the threshold $\tho$ during each active interval ${I_{A}}$ and ${I_{B}}$, so twice each, overall four threshold crossings. The \emph{segregation} percept is associated with responses where the overall number of threshold crossings of $u_A$ and $u_B$ per period is two, such that each tone's excitation crosses the threshold  $\tho$ only during its dominant active interval, as shown in Figure~\ref{po_pr_with_df}C. Observe that in Figure~\ref{po_pr_with_df}C $u_B(t)$ (red) stays below threshold  $\tho$ in $I_A$ (for $t<0.12$\,s) and $u_A(t)$ (blue) stays below threshold during $I_B$ (for $t>0.12$\,s). The \emph{bistability} percept is associated with responses where the total number of threshold crossings by $u_A(t)$ and $u_B(t)$ per period is three, such that the excitation of one tone crosses the threshold  $\tho$ during both active tone intervals, while the other does so only during its respective interval. Representative responses of bistable type are shown in Figure~\ref{po_pr_with_df}D (where $u_A(t)$ (blue) crosses once and $u_B(t)$ (red) crosses twice) and Figure~\ref{po_pr_with_df}E (where $u_B(t)$ (red) crosses once and $u_A(t)$ (blue) crosses twice).

For high repetition rates we also observe a \emph{saturated} regime without threshold crossings, where $u_A(t)$ and $u_B(t)$ stay above threshold $\tho=0.5$ for all time, which has not been reported in the experiments\cite{Van} (see Figure~\ref{fig:two_parameter_orgcase2}G).

\subsection{Perception boundaries - coherence and fission} 
  The boundary between two perception regimes is identified by excitation time profiles where $u_A(t)$ or $u_B(t)$ touch (or \emph{graze}) the threshold  $\tho$ during the active interval of the other tone (so $u_A(t)$ during $I_B$ and $u_B(t)$ during $I_A$). 

  For a symmetric time profile excitation $u_A(t)$ will graze the threshold  $\tho=0.5$ in $I_B$ if and only if $u_B(t)$ grazes the threshold during $I_A$. Thus,  symmetric time profiles with
\begin{align}
\begin{split}
\theta=\max\{ u_{B}(t):t\in[0,\Tinp/2]\}&=\max\{ u_{B}(t):t\in {I_A}\}\\
&= \max\{ u_{A}(t):t\in {I_B}\}
   \end{split} 
   \label{sym:criteria0}
\end{align}
will occur at the boundary between integrated and segregated perception region in parameter space.
An example excitation satisfying~\eqref{sym:criteria0} is shown in Figure~\ref{fig:two_parameter_orgcase2}B. 
For a non-symmetric time profile only one of the excitations grazes the threshold $\tho$ during the active interval of the other tone. Hence, we have four cases, coming in two symmetric pairs, resulting in two perception boundaries for non-symmetric responses.
At the \emph{coherence boundary} (between segregated and bistable perceptions) the response satisfies
\begin{align}
\begin{split}
  \tho&=\max\{ u_A(t):t\in[\Tinp/2,\Tinp]\}\\
  &=\max\{ u_A(t):t\in I_B\}, \ \mbox{where}\\
  %& \max\{ u_{A}(t):t\in[0,2\Tinp]\}=\max\{ u_{A}(t):t\in {I_B}\}=\theta, \ \mbox{with}\\
   \tho&>\max\{ u_B(t):t\in I_A\}.
   \end{split} \label{non-sym:criteria}
\end{align}
An example profile is shown in Figure~\ref{fig:two_parameter_orgcase2}C. Due to symmetry there will also be a grazing response with $\theta=\max\{u_B(t):t\in I_A\}$ and $\theta>\max\{u_A(t):t\in I_B\}$ for the same parameter values.
At the \emph{fission boundary} (between integrated and bistable perceptions)the response satisfies
\begin{align*}
  \tho&=\max\{ u_A(t):t\in[\Tinp/2,\Tinp]\}=\max\{ u_A(t):t\in I_B\},\\% \ \mbox{and}\\
  %& \max\{ u_{A}(t):t\in[0,2\Tinp]\}=\max\{ u_{A}(t):t\in {I_B}\}=\theta, \ \mbox{with}\\
   \tho&<\max\{ u_B(t):t\in I_A\}.
 %\label{non-sym:criteria:fission}
\end{align*}
An example profile is shown in Figure~\ref{fig:two_parameter_orgcase2}D. Again, due to symmetry there will also be a response with $\theta=\max\{u_B(t):t\in I_A\}$ and $\theta<\max\{u_A(t):t\in I_B\}$ for the same parameter values.

\paragraph{Tracking of perception boundaries in parameter space} A necessary condition for the touching of the threshold $\theta$ can be identified by introducing the time of grazing as a variable $t_\mathrm{thr}$, and adding two equations to the auditory model \eqref{model},\,\eqref{inpts_smooth_fun} %  (or the extended model \eqref{modelhopf} in Appendix~\ref{subsec1:audi:model})
\begin{align}
  u_{A}(t_\mathrm{thr})&=\theta,& u'_{A}(t_\mathrm{thr})&=0. \label{touchingr_eq}
\end{align}
We can track a branch of symmetric and a branch of nonsymmetric solutions of \eqref{model},\,\eqref{inpts_smooth_fun},\,\eqref{touchingr_eq}, for which $u_A$ touches the threshold $\theta$ by considering $t_\mathrm{thr}$ as an additional free parameter. The tracking of the integration-segregation boundary is initialized by a (symmetric) periodic orbit that satisfies the condition in \eqref{sym:criteria0} approximately. The tracking of the fission and of the coherence boundary is initialized by a non-symmetric periodic orbit satisfying \eqref{non-sym:criteria}.
\section{Results}\label{sec:bifanalysis}
In this section,  we provide a systematic bifurcation analysis of the auditory streaming model~\eqref{model} with forcing \eqref{inpts_smooth_fun}, where the two primary varied parameters are the presentation rate $\rp$ (inversely proportional to forcing period $\Tinp$) and the difference in tone frequency $\df$. Both parameters can be easily varied in experiments. As \texttt{DDE-Biftool}\cite{Jan,ELR02}, the numerical continuation software for DDEs we use, does not support problems that depend explicitly on time $t$, we convert the system into an autonomous system by appending a Hopf bifurcation normal form system, which we then use as a periodic driver to generate the periodic inputs $i_A$ and $i_B$. The complete extended system used to generate figures~\ref{po_pr_with_df}--\ref{fig:Cases_TD_TDD}  is given in Appendix~\ref{subsec1:audi:model} as system~\eqref{modelhopf}. By tracking the symmetry-breaking bifurcation branch  and families (branches) of stable periodic responses that satisfy grazing criterion \eqref{touchingr_eq} in the $(\rp,\df)$-plane, we initially check how the model reproduces the experimental diagram in Figure~\ref{Sigmoid_plot}. Then we will vary further parameters to make predictions how properties of the feedback loop in the secondary auditory cortex will influence experimental observations. 
\subsection{Bifurcation analysis}
Figure~\ref{po_pr_with_df}A shows a one-parameter branch of symmetric periodic orbits  and a branch of nonsymmetric periodic orbits, when varying the presentation rate $r_{p}$, keeping the frequency difference $\df$ fixed at $0.73$ and all other parameters as in Table~\ref{tab:parameters:exper}.

The $y$-axis in Figure~\ref{po_pr_with_df}A is a measure of non-symmetry. It takes the average difference between $u_B$, and $u_A$ (shifted by a half period) over one periodic orbit,  which is given by  the quantity: 
\begin{align}
    \mu_{\bot}=\frac{2}{\Tinp}\int_0^{\Tinp/2}u_{B}(t)-u_{A}(t+\Tinp/2)dt.\label{measure_symmetry}
\end{align}
This measure $\mu_{\bot}$ is zero for symmetric periodic orbits, and non-zero for non-symmetric periodic orbits.
% We fix the tone frequency difference at $\df=0.73$, maintaining all other parameters as specified in Table~\ref{tab:parameters:exper}.
The branch of symmetric periodic orbits undergoes symmetry-breaking bifurcations at $\rp= 3.44$ and $\rp = 7.47$ (black dots in Figure~\ref{po_pr_with_df}A). At these points a loss of stability occurs for the symmetric periodic orbit and two branches of non-symmetric periodic orbits emerge (which are symmetric images of each other). The purple solid line at $\mu_{\bot}=0$ in Figure~\ref{po_pr_with_df}A represents the branch of stable symmetric periodic orbits. The black dashed segment represents the unstable part of the symmetric branch. Starting from the detected symmetry-breaking point, we branch off toward the nonsymmetric periodic orbits (solid green curves in Figure~\ref{po_pr_with_df}A), which are all stable.

Based on the threshold criterion, the branch of stable symmetric periodic orbits corresponds to responses that give an integrated perception at low presentation rates and a segregated perception at high presentation rates. Along the branch of stable nonsymmetric periodic orbits the perception becomes bistable at the small blue circles in Figure~\ref{po_pr_with_df}A. Time profiles corresponding to each perception are shown in Figure~\ref{po_pr_with_df}B--E. The response at the red point  on the left stable  branch of symmetric periodic orbits shows integrated behavior of neural activity (Figure~\ref{po_pr_with_df}B), while the response at the brown point on the right stable branch of  periodic orbits shows segregated behavior (Figure~\ref{po_pr_with_df}C).  The responses at the red crosses on the branch of the nonsymmetric periodic orbits both show bistabile perception and are symmetric images of each other (Figure~\ref{po_pr_with_df}D and E).

\subsection{Perceptual organization and relative boundaries}
In order to find the boundaries between perceptual states, we continue the bifurcation analysis in the two-parameter plane $(\rp, \df)$, tracking the symmetry-breaking bifurcation points. We use the detected symmetry-breaking point at $\rp=3.44$ as starting point. The resulting symmetry-breaking branch is shown as a blue curve in the $(\rp, \df)$-plane in Figure~\ref{fig:two_parameter_orgcase2}A.

To identify the regions of parameter space according to their perceptual classification based on the threshold criterion, we compute branches of periodic orbits that graze the threshold at $\theta = 0.5$. Starting from a symmetric periodic orbit that grazes the threshold, and including equations~\eqref{touchingr_eq}, we track a family of symmetric periodic orbits satisfying condition~\eqref{sym:criteria0}. The resulting curve in the $(\rp,\df)$-plane is shown in Figure~\ref{fig:two_parameter_orgcase2}A, where stable symmetric grazing periodic orbits are represented by the pink solid branch segment, and unstable symmetric grazing periodic orbits are represented by the grey branch segment. An example time profile of a stable symmetric grazing response is shown in Figure~\ref{fig:two_parameter_orgcase2}B (parameters indicated by pink cross). Also using \eqref{touchingr_eq}, we track a branch of nonsymmetric periodic orbits that touch the threshold at $\theta = 0.5$, starting from a nonsymmetric periodic orbit that satisfies the nonsymmetric criterion  \eqref{non-sym:criteria}. The resulting curve is shown with black dashed lines in Figure~\ref{fig:two_parameter_orgcase2}A, consisting entirely of stable (nonsymmetric) responses. Two example time profiles are shown in Figure~\ref{fig:two_parameter_orgcase2}C and D (parameter values indicated by yellow and green cross in Figure~\ref{fig:two_parameter_orgcase2}A). The response at the lower $\df$ (green cross) is on the fission boundary, while the response for the high $\df$ (yellow cross) is on the coherence boundary. As the results show, there are two possible transitions between integrated and segregated perception: a direct transition at higher presentation rates (without symmetry-breaking; Figure~\ref{fig:two_parameter_orgcase2}B) and a two-step transition at lower rates, involving two symmetry-breaking bifurcations (Figure~\ref{fig:two_parameter_orgcase2}C,\,D).

The computational results show that the curve of symmetry-breaking bifurcation points and the perceptual boundary based on threshold criterion --- characterized by nonsymmetric solutions touching the threshold --- form distinct boundaries rather than coinciding. Since these two boundaries do not align, there can be bistability between non-symmetric solutions in the dynamical systems sense that might not result in bistable perceptions according to the threshold criterion \eqref{non-sym:criteria}. For example, this occurs for the periodic orbits marked by the red cross in Figure~\ref{fig:two_parameter_orgcase2}A, which is between the symmetry-breaking branch (blue curve) and the grazing curve for nonsymmetric periodic orbits (black dashed). The time profile of this solution indicates that both populations $u_A$ and $u_B$ exhibit neural activity above the threshold during both active tone intervals (Figure~\ref{fig:two_parameter_orgcase2}D). So, this solution is classified as integrated based on the threshold criterion despite having dynamical bistability. 

For low frequency differences ($\df$) and high presentation rates ($\rp$) greater than 30\,Hz, a saturated dynamical state emerges in which the activities of both units remain above threshold (Figure~\ref{fig:two_parameter_orgcase2}F). A small frequency difference $\df$ causes high amplitude inputs for $i_A(t)$ during $I_B$ and $i_B(t)$ during $I_A$, while a high repetition rate $\rp$ causes successive tone intervals to alternate too rapidly relative to the decay of the neuron populations’ activities. This saturated state does not correspond to an experimentally reported auditory streaming percept (such as integration, segregation, or bistability) since in experiments $\rp$ typically ranges between 5 and 20\,Hz in relevant experiments. The boundary of this region, shown as a dashed green line in Figure~\ref{fig:two_parameter_orgcase2}A, was computed by tracking the branch of symmetric periodic orbits that graze the threshold from above. Figure~\ref{fig:two_parameter_orgcase2}E shows the time profile of a solution marked by a black cross on this branch. 

\subsection{Perceptual classification; variable neural threshold}\label{sec:secondpeaks}
\begin{figure}[tbp]
    \centering
    \includegraphics[width=0.49\textwidth]{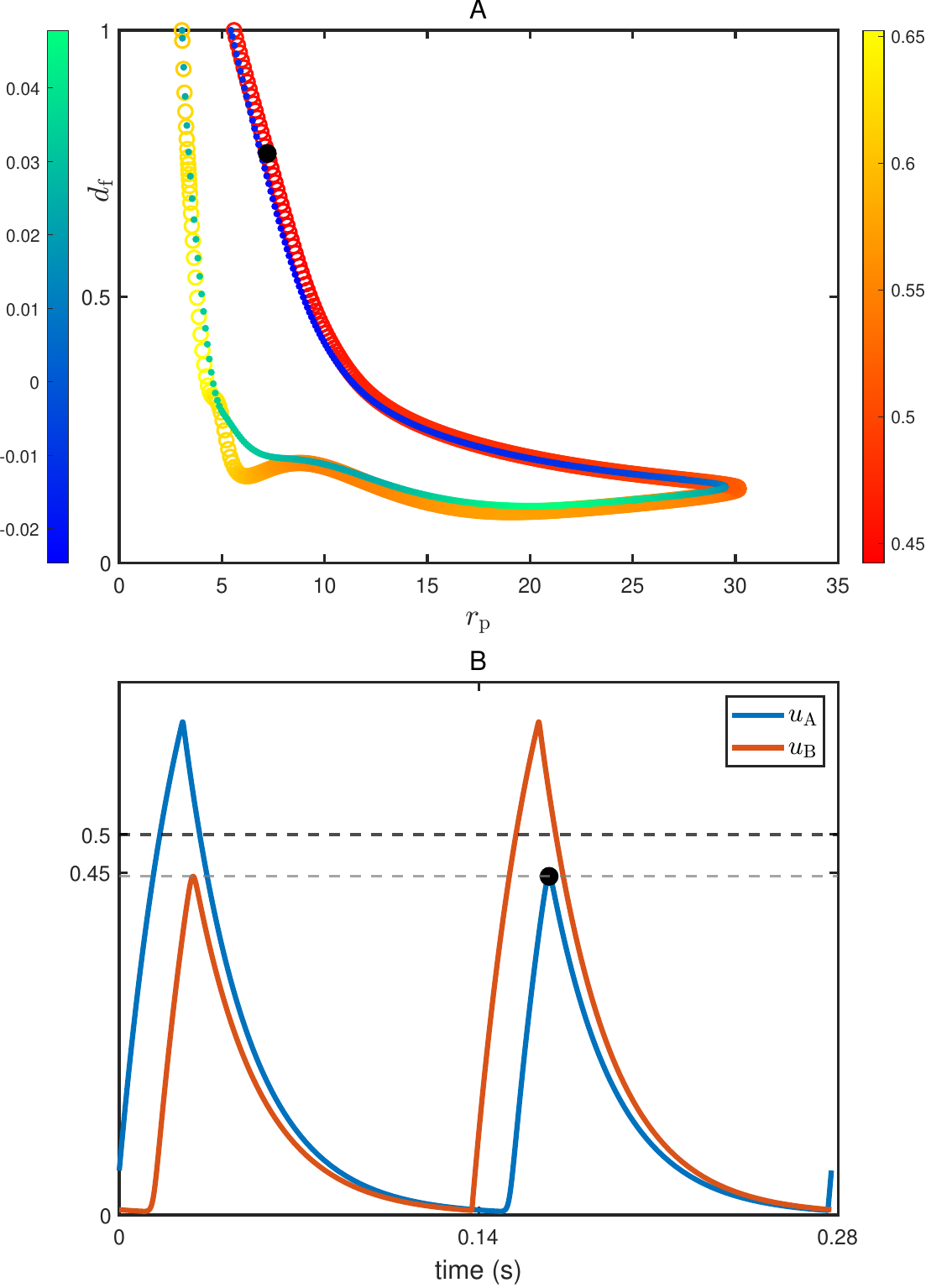}
    \caption{Distance between perceptual bistability and dynamical bifurcation. {\bf (A)}, the left colorbar (blue-to-green) represents the difference between peaks values of $u_A$ during $I_B$, and $u_B$ during $I_A$ per periodic forcing  along the non-symmetric branch with touching $\theta$ (perceptual boundaries). The right colorbar (dark to light orange and yellow) represents the values of the second peak for $u_{A}$ per periodic forcing along the symmetry-breaking curve.
     {\bf (B)} Time profile of the PO solution marked by a black dot on the symmetry-breaking branch in panel~A. The black dot in panel~B is an illustration of the second peak for $u_{A}$, along the symmetry-breaking branch. All other parameters are as specified in Table~\ref{tab:parameters:exper}.}
    \label{fig:secondpeaks}
\end{figure}

The results show that the curve of the dynamic symmetry-breaking bifurcation points and the perceptual boundary (based on the threshold criterion) nearly align. They also qualitatively align with the experimental map of perception boundaries in Figure~\ref{Sigmoid_plot}D in the reported range. The output of the secondary cortex will be processed by further activation, resulting in the reported perception. Our choice of threshold $\tho$ as $0.5$ implicitly assumes that this further activation's threshold equals $0.5$. To investigate how the location of threshold $\tho$ influences the distance between perception change and dynamical bifurcation, we analyze the secondary peak values of neural responses, associated with tone A ($u_{A}$) over tone B active interval (${I_{B}}$), for each periodic solution along the symmetry-breaking curve.  Figure~\ref{fig:secondpeaks}B shows an example time profile of a symmetric periodic response at symmetry breaking (parameter values indicated as black dot in Figure~\ref{fig:secondpeaks}A). The value of $u_A$ at its second peak (during interval $I_B$) equals approximately $0.45$ (grey horizontal dashed line). So, we report this value as a color code in Figure~\ref{fig:secondpeaks}A along the symmetry breaking curve in a yellow-to-red color coding. 
If the perception boundaries and the dynamical symmetry breaking  coincided, these secondary peaks would have to be equal to the threshold value of $\tho=0.5$ at the symmetry breaking bifurcation. Figure~\ref{fig:secondpeaks}A shows that the responses on the part of the symmetry breaking curve close to the fission boundary (integrated-to-bistable, lower in the parameter plane) has a secondary peak neural activity of approximately 0.55 to 0.65 (light orange and yellow color tones in Figure~\ref{fig:secondpeaks}A). The responses on the coherence boundary (segregated-to-bistable, upper part of curve in parameter plane) have a secondary peak neural activity ranging from approximately 0.45 to 0.55. (dark orange to red in Figure~\ref{fig:secondpeaks}A).

These results give an estimate how the perceptual boundary depends on the perception threshold and indicate how a dependence of threshold $\tho$ on signal parameters $\rp$ and $\df$ influences perception boundaries. As dynamical bistability is necessary for perceptual bistability  the varying secondary peaks along the symmetry breaking suggest that the auditory perceptual classification is based on a dynamic threshold (that is, one that depends on $\rp$ and $\df$). This adaptive threshold could be determined by analyzing the distribution of neural activity levels along the symmetry-breaking curve. 

Dynamical bistability is not sufficient for perceptual bistability because the contrast between secondary peak values is initially small near the (supercritical) symmetry breaking, such that further processing may not pick up the difference between the secondary peaks in $I_A$ versus $I_B$. The blue-green color coding along the threshold touching curve in Figure~\ref{fig:secondpeaks}A shows $\mu_\bot$ as defined in \eqref{measure_symmetry}, which is a measure for this contrast between secondary peaks.
A large $\mu_\bot$ in absolute value implies more certainty and less variability of the respective perception boundary. We observe that this is the case for the integrated-bistable (fission) boundary (lower part of symmetry breaking in $(\rp,\df)$-plane, about $4\%$). The low $\mu_\bot$ contrast of at most $2\%$ near segregated-bistable boundary implies that this perceptual boundary is more gradual.
\begin{figure*}[tbp]
    \centering
    \includegraphics[width=0.85\textwidth]{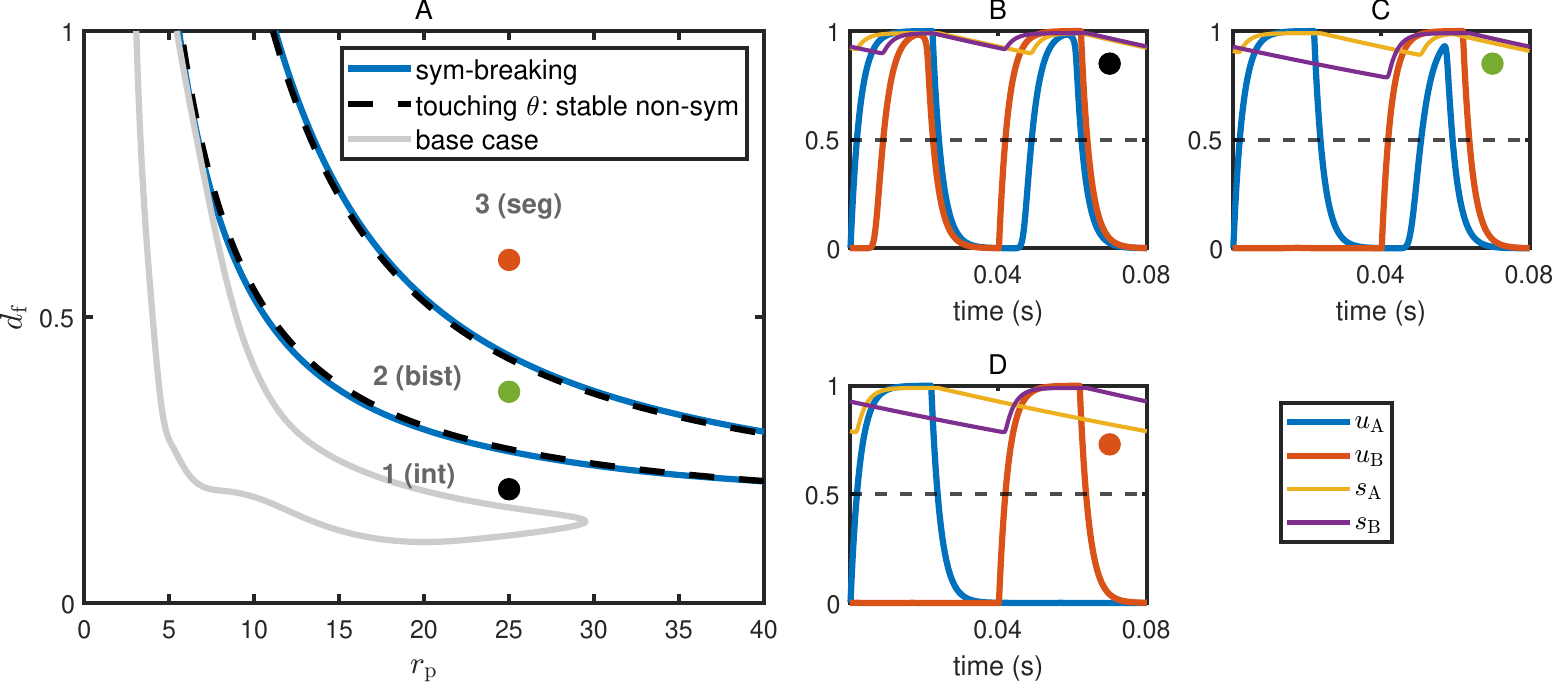}%results_small_tau2v5.pdf}%results_small_tau2.pdf}%smalltau_twoparameterresults.pdf}
    \caption{{\bf (A)} Two-parameter bifurcation analysis for the auditory model with a small time scale $\tau=0.0025$. Compared to the previous case ($\tau=0.025$; grey curve), the curves of the symmetry-breaking bifurcation point (blue curve) and the perceptual boundary (black dashed curve; threshold criterion) overlap and shift to the left and upward. {\bf (B-D)} Time profiles of the POs highlighted by the black, green, and red dots in panel~A, representing the integrated, bistability, and segregated, respectively. All other parameters are as specified in Table~\ref{tab:parameters:exper}.}
    \label{pulses_timeprofile}
\end{figure*}
\subsection{Auditory streaming with faster time scale for excitation}\label{sec:fastTimescale}
We observe that many time profiles in figures \ref{po_pr_with_df}B--E, \ref{fig:two_parameter_orgcase2}B--F and \ref{fig:secondpeaks}B do not show an approximate all-or-nothing excitatory response, but rather a brief crossing of the threshold over a narrow time range. Intuitively one expects that the perception boundary becomes less sensitive to the choice of threshold if the response ($u_A(t)$ and $u_B(t)$) more closely resembles an excitatory all-or-nothing response. One modification of the model leading to more prevalent all-or-nothing excitatory response is to consider a smaller time scale for neural excitation, $\tau$ ($\tau \ll 1$), while keeping the other parameters as in Table~\ref{tab:parameters:exper}. For Figure~\ref{pulses_timeprofile} we modified the time scale parameter by setting $\tau=0.0025$ (which is, however, unrealistically fast, compare to the original realistic value $\tau=0.025$ in Table~\ref{tab:parameters:exper}), and repeated the bifurcation analysis in the $(\rp, \df)$-plane, tracking the branch of symmetry-breaking bifurcation (blue curve) and the branch of nonsymmetric responses touching the threshold at $\theta = 0.5$ (black dashed curve) as illustrated in Figure~\ref{pulses_timeprofile}A (reference from bifurcation Figure~\ref{fig:two_parameter_orgcase2} shown in grey). The symmetry-breaking bifurcation and the perceptual boundary (threshold criterion) now align tightly, but they also exhibit a leftward and upward shift (which can be compensated by increasing $\thi$). This adjustment in the time scale leads to more realistic all-or-nothing neural pulses, where the values of $u_{A}$ and $u_{B}$ either reach levels close to unity (thus, exceeding the threshold $\tho$ significantly) or remain close to zero (see Figure~\ref{pulses_timeprofile}B--D).

\subsection{Effects of further parameters: tone duration $(t_{d})$ and inhibitory delay $(D)$}
\begin{figure*}[tbp]
    \centering
    \includegraphics[width=\textwidth]{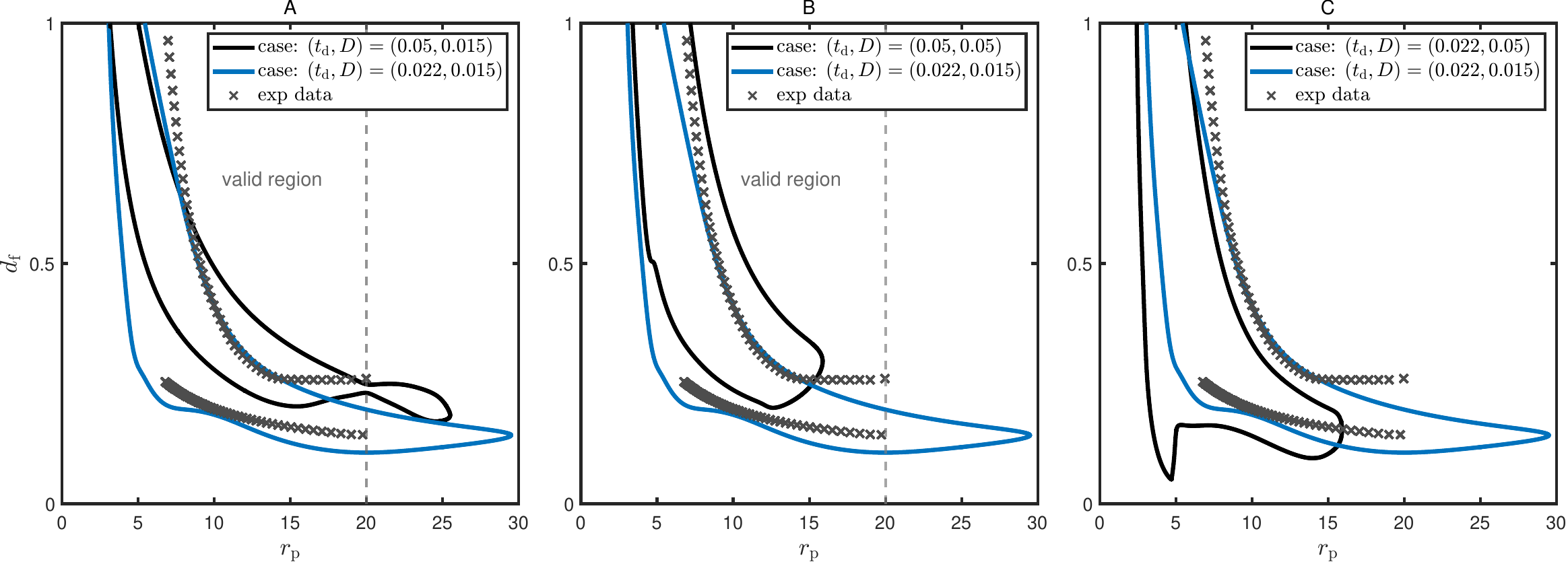}
    \caption{Two-parameter bifurcation diagrams for combinations of delays $D$ and tone duration $t_d$. The black curve represents nonsymmetric POs that touch the threshold, which represents the switching boundary between different auditory perceptions for different delay $D$ and tone duration $t_{d}$ parameter values. {\bf (A)} First case, increasing $t_{d}$ to 0.05.  {\bf (B)} Second case, increasing both $t_{d}$ and $D$ to 0.05. {\bf (C)} Third case: increasing $D$ to 0.05. The blue curve represents the perceptual boundary for the base case, as illustrated in Figure~\ref{fig:two_parameter_orgcase2}. The vertical dashed line at $r_p$  indicates the boundary of the valid region beyond which overlap occurs.}
    \label{fig:Cases_TD_TDD2}
\end{figure*}
\begin{figure*}[htbp]
    \centering
    \includegraphics[width=0.8\textwidth]{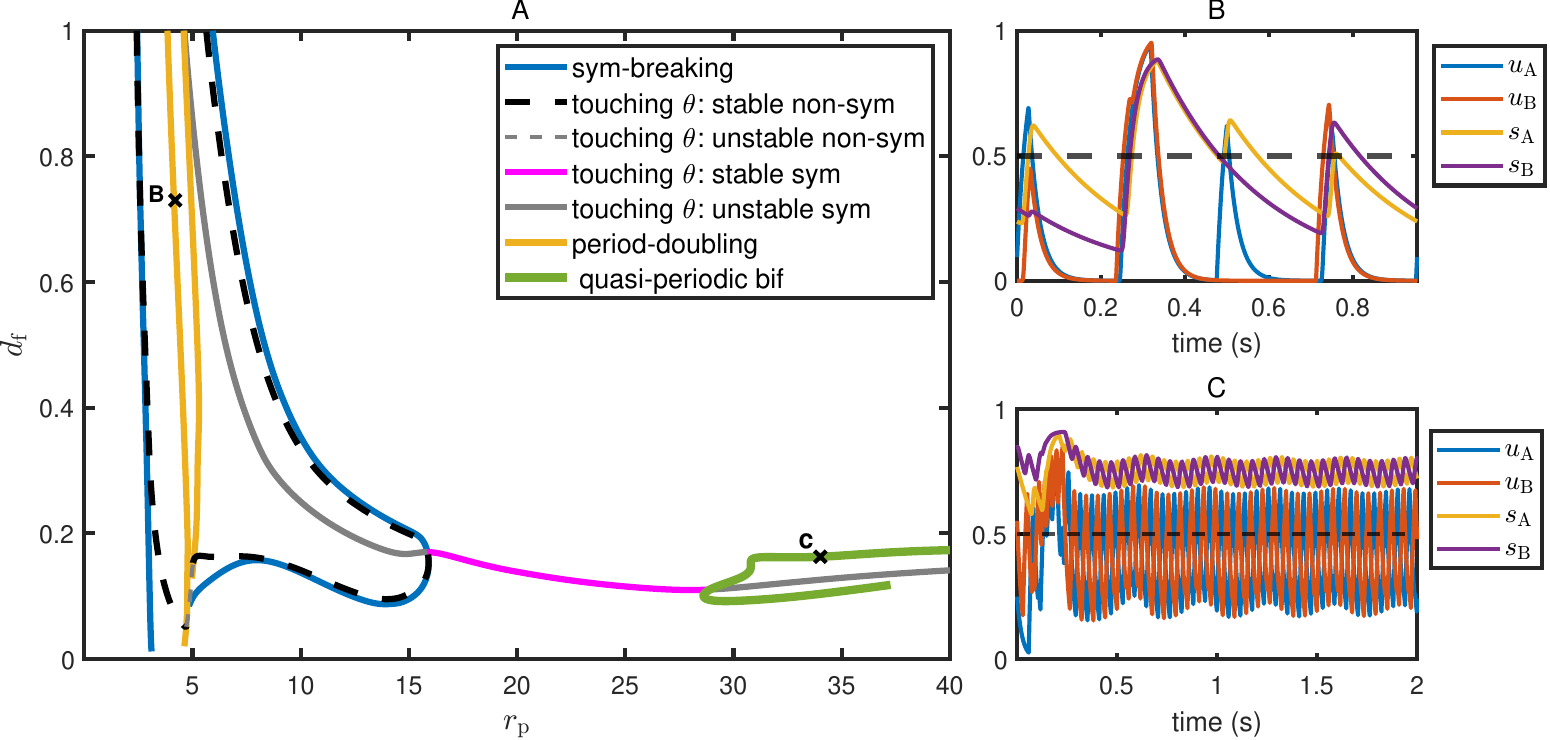}
    \caption{{\bf (A)} Detailed bifurcation analysis for the auditory model under the variation of $\rp$ and $\df$, for the case when increasing delay to 0.05, $(t_{d}, D)=(0.022,0.05)$. Other parameters are as specified in Table~\ref{tab:parameters:exper}. {\bf (B)} Time profile of a stable periodic orbit on the period-doubling branch, shown by yellow curve in panel~A. {\bf (C)} Quasi-periodic dynamics computed with initial conditions selected near the torus bifurcation branch shown by green curve in panel~A.}
    \label{fig:Cases_TD_TDD}
\end{figure*}

Finally, Figure~\ref{fig:Cases_TD_TDD2} shows how tone duration $t_{d}$ and inhibitory delay $D$ affect perceptual boundaries. In particular, we consider three cases for these parameter values:
\begin{itemize}
    \item (Figure~\ref{fig:Cases_TD_TDD2}A) increasing duration $t_{d}$ to $0.05$ keeping delay $D$ fixed at $0.015$, 
    \item (Figure~\ref{fig:Cases_TD_TDD2}B)  increasing both, $t_{d}$ and $D$, to $0.05$, and
    \item (Figure~\ref{fig:Cases_TD_TDD2}C) increasing $D$ to $0.05$ keeping $t_{d}$ fixed at $0.022$. The bifurcation diagram in Figure~\ref{fig:Cases_TD_TDD} shows more details of the dynamics for this case. 
\end{itemize}
In Figures~\ref{fig:Cases_TD_TDD2}A--C black  solid curves are the loci where the nonsymmetric responses touch the threshold $\tho=0.5$ for the respective value pair of  delay $D$ and tone duration $t_{d}$. Each panel includes the perceptual boundary for bistability for the base case from Figure~\ref{fig:two_parameter_orgcase2} (blue curves in Figures~\ref{fig:Cases_TD_TDD2}A--C).

Increasing tone duration $t_{d}$ to $0.05$ (Figure~\ref{fig:Cases_TD_TDD2}A) shifts the bistability region upward in tone frequency difference $\df$ (lower bound $0.17$ now) and reduces its range in presentation rate $\rp$ to $[3.15, 25.6]$\,Hz. For this tone duration parameter value, the presentation rate $\rp$ cannot exceed 20\,Hz, as beyond this value, the active tone intervals $I_A$ for tone A and $I_B$ for tone B would overlap. The vertical dashed line at $\rp$  indicates this upper boundary of the region of physically meaningful parameters $\rp$. 

When both, tone duration $t_d$ and delay $D$, are increased to $0.05$  (Figure~\ref{fig:Cases_TD_TDD2}B), the bistability region is confined mainly to a smaller range $\rp \in [3.2, 16]$\,Hz for $\df \in [0.15, 1]$, with a slight expansion to $\df \in [0, 1]$ for $\rp \in [4.08, 4.34]$.

Increasing the delay $D$ to $0.05$ but keeping $t_d=0.022$ (Figure~\ref{fig:Cases_TD_TDD2}C) shifts the bistability region downward in $\df$ and confines it to $\rp \in [2.45, 15.9]$ for $\df \in [0, 1]$.

Figure~\ref{fig:Cases_TD_TDD}A shows that for larger delay $D=0.05$ additional bifurcations occur: for example, a period-doubling bifurcation of nonsymmetric responses (yellow in FIgure~\ref{fig:Cases_TD_TDD}A) was observed, indicating the presence of period-two responses.  A time profile of a stable period-two response is shown in Figure~\ref{fig:Cases_TD_TDD}B. Additionally, for large $D=0.05$, a torus bifurcation was detected for symmetric responses at $\rp \approx 29$ (green curve in Figure~\ref{fig:Cases_TD_TDD}A), marking the onset of quasi-periodic behavior for high presentation rate regimes.Figure~\ref{fig:Cases_TD_TDD}C presents a time series for a solution for parameters near torus bifurcation curve, showing quasi-periodic behavior.

\section{Discussion}\label{sec:discussion}
%\subsection{Summary}
Our bifurcation analysis of a mathematical model of auditory streaming of two interleaved tone sequences with delayed cross-inhibition coupling\cite{J} traces the precise boundaries between different dynamical regimes in the two main experimental control parameters, the presentation rate $\rp$ and the frequency difference $\df$. It also determines how these boundaries depend on other not directly controllable parameters: the tone duration $t_{d}$, the delay $D$ in inhibitory coupling, and the time scale difference $\tau$ between the fast and slow dynamics. There are two implications for experiments , (1) the hypothesis that the neural threshold for perceptual boundaries may be dynamic, and (2), the effect of internal parameters on the perceptual bistability region, which may allow one to draw conclusions about these internal not directly accessible parameters.

\subsection{Variable neural threshold on perceptual boundary}
Our results in Figure~\ref{fig:secondpeaks} suggest that the threshold values for fission and coherence boundaries appear to be variable, rather than occurring at a fixed threshold. Specifically, the threshold for the fission boundary tends to be higher than the threshold for the coherence boundary. Physiological evidence from neural activity recordings in the primary auditory cortex of awake monkeys supports these findings by comparing differential attenuations across varying stimulus conditions with perceptual outcomes observed in human listeners~\cite{13}. The comparison suggests that smaller attenuations usually occur when stimulus conditions fall within the integrated perceptual region, while larger attenuations typically occur within the segregated perceptual region.

A uniform threshold $\theta$ for all experimental stimuli assumes a uniform neural criterion for distinguishing perceptual states, potentially oversimplifying the complex interplay between neural dynamics and perceptual outcomes. By contrast, a variable threshold reflects the context-dependent nature of neural processing, capturing how the auditory system adapts to changing stimulus features and internal states. This flexible classification approach could provide insights into how the brain resolves ambiguous auditory stimuli, supporting dynamic shifts between integration and segregation. Furthermore, variable thresholds could serve as a basis for refining computational models of auditory processing, enabling more accurate predictions of perceptual outcomes in realistic listening environments. Future experimental research investigations on whether the threshold should be different at these two boundaries and whether it should be fixed or variable depending on $\rp$ would clarify the nature of these perception thresholds.

\subsection{Effect of stimulus features}
Our mathematical framework enables the prediction of how model parameters and stimulus characteristics (e.g., inhibition coupling delay $D$, tone duration $t_d$, and input strengths) influence perceptual boundaries. The role of tone duration $t_d$ has not been extensively investigated experimentally and it is unclear if the duration $t_d$ as received from the primary cortex is proportional to the duration of the signal.  Studies of human listeners exposed to galloping stimulus sequences (ABA-ABA-\ldots) suggest that increasing tone duration favors segregated perception (reported difficulty perceiving the gallop)% the differential attenuation of B-tone responses
~\cite{bregman2000effects, fishman2001neural}. Our results indicate that the relationship between tone duration and stream segregation is more nuanced. Specifically, our model predicts that at lower presentation rates, increasing tone duration promotes segregation, whereas at higher presentation rates, it inhibits segregation. Furthermore, the shape of the region of perceptual bistability in the $(\rp,\df)$-plane permits some conclusions on the coupling delay $D$: a larger delay $D$ reduces bistability at higher presentation rates.

\subsection{Limitations and future work}
Future work will involve tuning the model parameters to align the dynamically bistable region with experimental data more precisely, particularly the coherence and fission boundaries that classify auditory percepts into integration, segregation, and bistability. A key limitation of the initial bifurcation analysis is the mismatch between the symmetry-breaking curve and the switching boundary between two different auditory perceptions, represented by the nonsymmetric curve of POs touching the threshold (see Figure~\ref{fig:two_parameter_orgcase2}A). To address this, we compare with a smaller time scale for neural excitation ($\tau=0.0025$), which resulted in the coincidence of the symmetry-breaking curve and the nonsymmetric curve of POs touching the threshold. Despite these improvements, this adjustment introduced a misalignment between the perceptual boundaries and the experimental data~\cite{46} (see Figure~\ref{pulses_timeprofile}A). To resolve this issue, future work should focus on refining the model to better align with experimental data while preserving the correspondence between mathematical and perceptual bistability. One promising approach is to explore larger values of the delay parameter $D$. Increasing the delay shifts the bifurcation curves to the left, counteracting the shift of perception boundaries caused by the larger time-scale difference $\tau$. This adjustment is expected to improve the agreement between the model and the experimental coherence and fission boundaries. 

Furthermore, an important aspect not addressed in the current analysis is the influence of noise on auditory perception. In neural systems, noise plays a role in shaping perceptual outcomes, particularly in perceptual bistability, where slight variations can lead to significant changes in perceptual states~\cite{shpiro2009balance,moreno2007noise}. Introducing noise into the input (thus, the parameters) or into the model permits studying how stochastic perturbations interact with the bifurcation structure, generating spontaneous (noise-induced) transitions between different auditory perceptions. For example, noise can induce transitions between perceptual states in regions where the system is otherwise stable or amplify small differences in input signals, leading to more variable behavior~\cite{darki2020methods,darki2023hierarchical}. Future work should thus focus on systematically characterizing the effects of different types and intensities of noise on the dynamics of auditory bistability.

\subsection{Acknowledgements}
We thank James Rankin and Andrea Ferrario for their valuable discussions and insightful suggestions during this project.

\subsection{Availability of data and materials}
Source code for the model will be available in a GitHub repository
\begin{center}
  {\small{\urlstyle{tt}\url{github.com/asim-alawfi/auditory-model-publication}}}.
\end{center}

\subsection{Funding}
AA gratefully acknowledges the support for his scholarship at the University of Exeter, provided by Imam Mohammad Ibn Saud Islamic University (IMSIU). FD acknowledges support from an Engineering and Physical Sciences Research Council (EPSRC) Standard Grant (Healthcare Technologies), (EP/W032422/1). 

For the purpose of open access, the corresponding author has applied a 'Creative Commons Attribution' (CC BY) licence to any Author Accepted Manuscript version arising from this submission.

\subsection{Competing interests}
The authors declare that they have no competing interests.

\subsection{Authors’ contributions}
All authors were involved in the problem formulation and discussion of the results. JS and FD jointly supervised the work. All authors contributed to the manuscript. AA implemented and carried out all numerical analyses and simulations. All authors read and approved the final manuscript.

\appendix
\section{Extension of auditory model to autonomous system} \label{subsec1:audi:model}
Section~\ref{sec:bifanalysis} performs bifurcation analysis of \eqref{model} with forcing \eqref{inpts_smooth_fun}. As \texttt{DDE-Biftool}\cite{Jan,ELR02}, the numerical continuation software for DDEs we use, does not directly support problems that depend explicitly on time $t$, we need to convert the system into an autonomous system. This can be achieved by extending \eqref{model} with a planar oscillator that generates a stable sinusoidal limit cycle (e.g., Hopf normal form equations\cite{ku}), and then replacing $\sin(t)$ by its first component in \eqref{inpts_smooth_fun}. The resulting extended autonomous system has the form (dropping the argument $t$ from instantaneous terms)
\begin{equation} 
\begin{split}
\tau\dot{u}_{A}&=-u_{A}+\sig_{\theta}\left(au_{B}-bs_{B}(t-D)+i_{A}\right), \\
\tau\dot{u}_{B}&=-u_{B}+\sig_{\theta}\left(au_{A}-bs_{A}(t-D)+i_{B})\right),\\
\tau\dot{s}_{A}&=-(\tau/\tau_{i})\,s_{A}+\sig_{\theta}(u_{A}\left(1-s_{A}\right) , \ \ \ \ \ \\
\tau\dot{s}_{B}&=- (\tau/\tau_{i})\, s_{B}+\sig_{\theta}(u_{B})\left(1-s_{B}\right), \\
\dot{y}_1&=\alpha y_1 - \omega y_2-y_1\left(y^{2}_1+y^{2}_2\right),\\
\dot{y}_2&=\omega y_1 + \alpha y_2-y_2\left(y^{2}_1+y^{2}_2\right),
\end{split} \label{modelhopf}
\end{equation}
where $\alpha=1$, $\omega=\pi\rp$, and
\begin{equation}
\begin{aligned} 
    i_\text{A}&=c\sig_{0}(y_{1}) \sig_{0}(-y_{1}(t-t_d))+d\sig_{0}(-y_{1})\sig_{0}(y_{1}(t-t_d)),\\
    i_\text{B}&=c\sig_{0}(-y_{1})\sig_{0}(y_{1}(t-t_d))+d\sig_{0}(y_{1}) \sig_{0}(-y_{1}(t-t_d)),
    \end{aligned}\label{force_hopf}
\end{equation}
The differential equations for $y_{1,2}$ have a stable harmonic limit cycle, which implies that $y_1(t)=\sin(\omega t)$ if $y_1(0)=0$.\\

%\bibliography{aipsamp}% Produces the bibliography via BibTeX.

\end{document}